\documentclass[superscriptaddress,nobibnotes,amsmath,amssymb,notitlepage,twocolumn,prl,longbibliography]{revtex4-1}

\usepackage{bm,mathptmx,braket}

\usepackage{graphicx,color,hyperref}
\usepackage[caption=false]{subfig}
\hypersetup{colorlinks=true, linkcolor=blue, citecolor=blue, urlcolor=blue} 

\graphicspath{{./img/}}

\newcommand{\beq}[1]{\begin{equation}\label{#1}}
\newcommand{\eep}{\;.\end{equation}}
\newcommand{\eec}{\;,\end{equation}}
\newcommand{\eeq}{\end{equation}}

\newcommand*\dd{\mathop{}\!\mathrm{d}} 

\newcommand{\om}{\omega}

\newcommand{\sect}[1]{\vspace{0.3em}{\it #1.}---}

\DeclareMathAlphabet{\mathcal}{OMS}{cmsy}{m}{n} 

\renewcommand{\vec}[1]{{\bf #1}}

\newcommand{\kv}{\vec{k}}

\begin{document}

\title{Topological Optical Chirality Dichroism}

\newcommand{\UoM}{{Department of Physics and Astronomy, University of Manchester, Oxford Road, Manchester M13 9PL, United Kingdom}}

\newcommand{\TCM}{{Theory of Condensed Matter Group, Cavendish Laboratory, University of Cambridge, J.\,J.\,Thomson Avenue, Cambridge CB3 0HE, United Kingdom}}

\newcommand{\Coimbra}{{Department of Physics, University of Coimbra, Rua Larga, 3004-516 Coimbra, Portugal}}

\newcommand{\Dublin}{{School of Theoretical Physics, Dublin Institute for Advanced Studies,
10 Burlington Road, Dublin D04 C932, Ireland}}


\author{Wojciech J. Jankowski}
\email{wjj25@cam.ac.uk}
\affiliation{\TCM}

\author{Giandomenico Palumbo}
\email{giandomenico.palumbo@gmail.com}
\thanks{These authors contributed equally.}
\affiliation{\Coimbra}
\affiliation{\Dublin}

\author{Robert-Jan Slager}
\email{robert-jan.slager@manchester.ac.uk}
\thanks{These authors contributed equally.}
\affiliation{\UoM}
\affiliation{\TCM}

\date{\today}

\begin{abstract}
We report on a universal topological dichroism of chiral three-dimensional systems in response to the chirality of light. We show that chiral topological invariants result in integer-quantized dichroic excitation rate differences. Moreover, we demonstrate that such topological effects arise more generally from coupling optical chirality to higher tensor Berry curvatures and Dixmier--Douady invariants of quantum states, including Hopf indices. We finally propose an experimental setup that leverages superchiral light as a smoking-gun probe of chiral band topologies in three-dimensional materials. Our findings establish an optical route for probing to date unobserved chiral electronic band topologies.
\end{abstract}

\maketitle
\sect{Introduction} Quantization and topological phenomena constitute a modern cornerstone in physics. Fueled by the discovery of quantum Hall effects~\cite{Tsui1982, Laughlin1983, vonKlitzing1986}, an ongoing characterization of phases with topological invariants, rather than by symmetry-breaking in order parameters, has transformed the understanding of quantum matter. This wide and active research scene has been put on a firm footing upon elucidating the precise mathematical nature of the quantized Hall response~\cite{Simon1983} in terms of so-called Chern invariants~\cite{Chern1974}, and by making direct connections with Chern-Simons field theories originally explored in high energy contexts~\cite{Witten1989, Witten1995}. The reach of these pursuits is further motivated by their pioneering role in the discovery of topological insulators and semimetals~\cite{Rmp1,Rmp2,Rmp3}, whose development followed a reverse path in time. That is, after predictions of the possibility of refined topological invariants in the presence of nonspatial~\cite{Kitaev2009,SchnyderClass} and spatial symmetries~\cite{kruthoff2017,Bradlyn2017,Po2017}, topological materials were experimentally observed~\cite{Hsieh2008}, and the corresponding field theories were formulated~\cite{Qi2008}. Currently, all these research activities are still in full force and encompass searches carried out in the range from the discovery of novel types of topological insulators and semimetals~\cite{Davoyan2024, Jankowski2024PRB}, to the finding of their possible response signatures in observables~\cite{Jankowski2025opticalprb, Jankowski2024PRL, Jankowski2025gerbe}. Given these immense advancements, the question whether topological effects akin to the original quantum Hall exist is a prominent, albeit a~challenging one, to ask. 
\begin{figure}[t!]
    \centering
    \includegraphics[width=\linewidth]{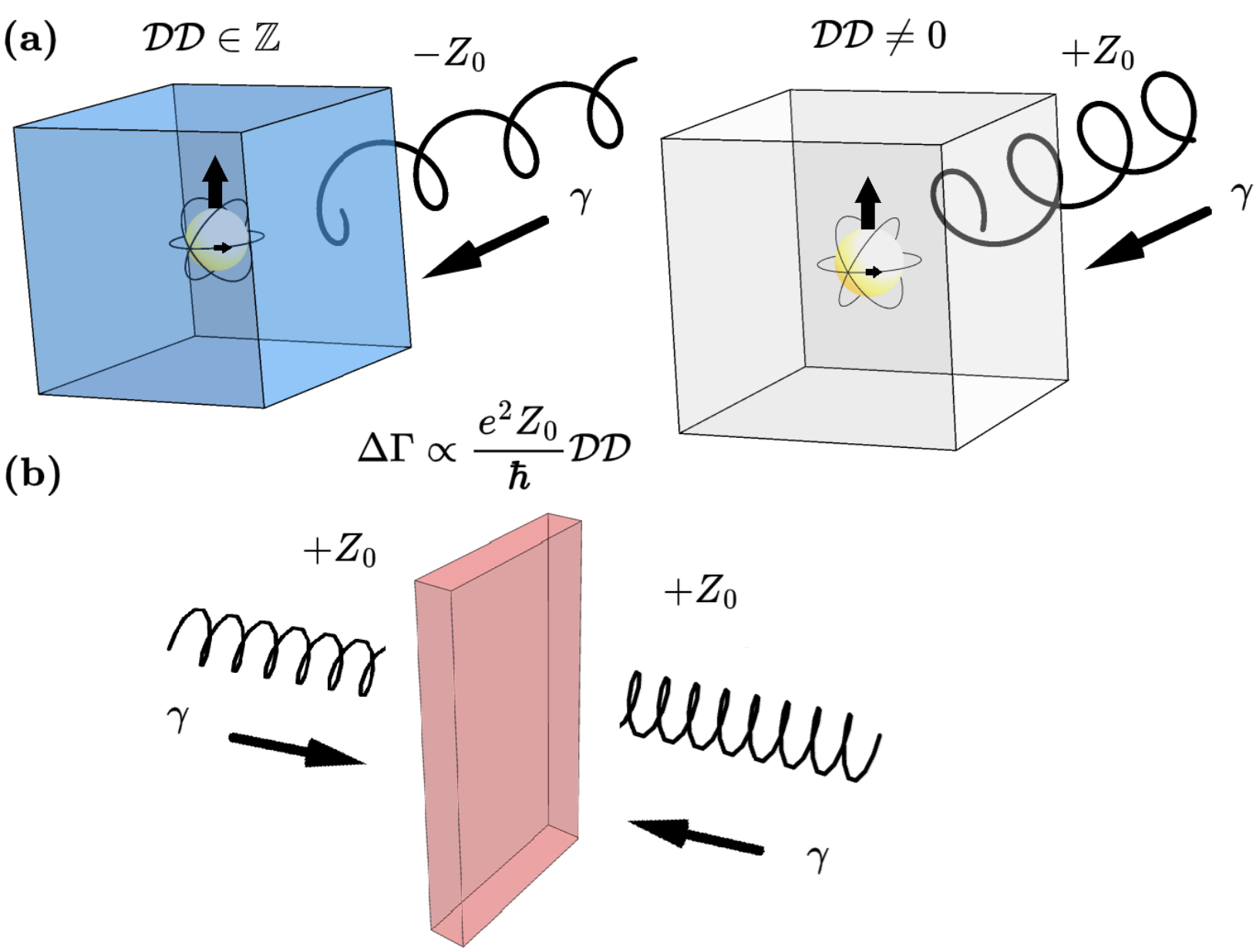}
    \caption{Topological optical chirality dichroism (TOCD). {\bf (a)}~\mbox{Chiral} topological insulator ($\mathcal{DD} \neq 0$) exhibiting dichroism between light beams ($\gamma$) with optical chiralities $\pm Z_0$, absorbing zilch $-Z_0$ (\textit{blue}), transmitting zilch $Z_0$ (\textit{white}). The effect arises from gyrotropic birefringence of chiral bulk electric currents supported by topologically nontrivial electrons. {\bf (b)}  Experimentally measurable absorptive chiral topological response to superchiral light with enhanced zilch $Z_0$. We establish TOCD of superchiral light as a~smoking-gun signature of chiral bulk topologies.}
    \label{Fig:Intro}
\end{figure}

In this Letter, we report on a surprising finding of such an effect, a new type of topologically quantized dichroism in response of three-dimensional (3D) chiral topological insulators to the optical chirality of light. We term the new effect as topological optical chirality dichroism (TOCD) and we relate it to the less known concept of zilch.
Originally introduced by David~M.~\mbox{Lipkin}~\cite{Lipkin1964} in the 60's, 
zilches represent higher-order conserved quantities of the electromagnetic field that are distinct from energy, momentum, and helicity \cite{Cameron_2012,Bliokh_2013}. Remarkably, recent work has shown that zilch is not merely a~formal construct, but can manifest itself in transport phenomena analogous to anomaly-induced responses~\cite{Chernodub2018}. In particular, zilch currents can be generated in rotating photon systems, giving rise to a vortical effect closely resembling chiral transport in relativistic fluids, and have been linked to gravitational anomaly–like contributions. Furthermore, a kinetic theory description has revealed that zilch transport is governed by Berry phase of photons, thereby establishing a direct connection between optical chirality and geometric phases \cite{Huang2020}. These results suggest that zilch provides a bridge between electromagnetic chirality and topological response theory. In \mbox{parallel}, nontrivial field configurations such as Hopfions \cite{Ranada,Kedia2013,Palumbo2025} have been shown to offer complementary realizations where chirality acquires a global topological character \cite{Smith_2018}. Taken together, these insights motivate viewing optical chirality as a~probe of both geometric and topological structures, providing a~natural conceptual underpinning for the emergence of TOCD introduced in this~work.

Apart from retrieving the first-of-a-kind universal quantized response to zilch, on a more fundamental side, we further elucidate that the TOCD roots in deeper connections with other intricate mathematical objects known as bundle gerbes~\cite{Murray1996, Jankowski2025gerbe, Bzdusek2025}. Previously, bundle gerbes were explored in string theories~\cite{Hitchin1999,Ekstrand2000,Szabo2017}, tensor gauge theories \cite{Carey2000,Borsten2025, Palumbo2025D}, topological band theory \cite{Palumbo2018,Palumbo2019,Zhu2020,Palumbo2021,Mo2022} and tensor network contexts~\cite{Ohyama2024,Ohyama2024_2,Sommer2025,Ohyama2025}, yet, their role in optics remained underexplored. 

We demonstrate that TOCD is supported by a broad class of topological insulators with nontrivial bundle gerbe invariants such as Dixmier--Douady classes that result in integer chiral winding numbers and Hopf indices~\cite{Jankowski2025gerbe}. Most importantly, the central focus of this work is on three-dimensional chiral topological insulators in symmetry class AIII \cite{Hosur2010}, which to the best of our knowledge lack a bulk optical signature of nontrivial band topology to date. These systems furnish a minimal 3D setting for a strong integer topological invariant protected solely by chiral symmetry, in contrast to chiral-invariant topological superconducting phases whose description intrinsically involves particle-hole symmetry through Bogoliubov-de Gennes Hamiltonians \cite{SchnyderClass}. Their integer winding number admits a clear bulk-boundary correspondence manifested as protected surface Dirac cones, making class AIII a natural platform for generalizing quantum-Hall-type topological physics to three dimensions. In addition, their experimental accessibility in synthetic-matter setups~\cite{Ji2020,Xin2020} makes them well-suited for a concrete investigation of TOCD in realistic 3D topological systems. We numerically show how TOCD not only allows one to optically probe stable class AIII topologies, which we demonstrate using established three-band and four-band models~\cite{Neupert2012, Palumbo2019, Liu2023}, but also how it emerges in chiral 3D phases with delicate Hopf indices~\cite{Moore2008, Kennedy2016, Alexandradinata2021, Nelson2022, Bzdusek2025}.
Before concluding, we propose how TOCD could be experimentally accessed with superchiral light, thereby opening up for unprecedented optical search pursuits for uncharted integer chiral topological invariants in 3D quantum materials.

\sect{Topological response to optical chirality} We start by unraveling the topological response to optical chirality. To this end, we recall that the optical chirality of light can be captured with a single pseudoscalar quantity~\cite{Tang2010},
\beq{}
    Z_0 = \vec{E} \cdot (\nabla \times \vec{E}) + \vec{B} \cdot (\nabla \times \vec{B}),
\eeq
the so-called zilch~\cite{Lipkin1964}. The chirality of light ($Z_0 \neq 0$) requires elliptical or circular light polarizations, while linearly-polarized light is not optically chiral $(Z_0 = 0)$, see End Matter. The magnitude of $Z_0$ can be tuned independently of light's energy density $u \sim \frac{1}{2}|\textbf{E}|^2$~\cite{Tang2010}, with $|Z_0|/u \gg 2|\textbf{q}|$ condition being definitional to superchiral light, where $\textbf{q}$ is the photon wavevector. The superchiral light can be engineered with counterpropagating beam alignments [Fig.~\ref{Fig:Intro}(b)], which we propose for probing chiral topologies considered in this work. In what follows, we demonstrate how the chirality of light ($Z_0$) allows to distinguish distinct chiral band topologies.

Central to this work, topological three-dimensional chiral matter can realize integer topological invariants, formally corresponding to the Dixmier--Douady ($\mathcal{DD}$) characteristic classes \cite{Palumbo2018,Palumbo2019,Zhu2020},
\beq{}
    \mathcal{DD} = \frac{1}{4\pi^2} \int_\text{BZ} \dd^3 \kv~ \mathcal{H}_{xyz} \in \mathbb{Z}.
\eeq

Here, $\mathcal{H}_{xyz} = \partial_{k_x} B_{yz} + \partial_{k_y} B_{zx} + \partial_{k_z} B_{xy}$ is the torsion flux of tensor Berry connection $B_{ij}$~($i,j=x,y,z$) over the space of single-particle momenta of electrons $\textbf{k} = (k_x, k_y, k_z)$ in the first Brillouin zone (BZ)~\cite{Jankowski2025gerbe}. The tensor Berry connection ${B_{ij} = \phi F_{ij}}$, resulting in torsion $\mathcal{H}_{xyz}$, arises from the standard non-Abelian Berry curvature $F_{ij}$ gauged with a pseudoscalar $\phi$~\cite{Palumbo2021}, following the construction proposed in Ref.~\cite{Jankowski2025gerbe}, which we also detail in the Supplemental Material (SM)~\cite{SI}.

We note that $\mathcal{DD} = 0$ in presence of mirror symmetries ($\mathcal{M}$), as the three-form $\mathcal{H}_{xyz}$ satisfies $\mathcal{M}\mathcal{H}_{xyz}\mathcal{M}^{-1} = -\mathcal{H}_{xyz}$. This transformation property fundamentally reflects the intrinsic chirality of electronic orders in any topological insulators with $\mathcal{DD} \neq 0$. As a result, we find that the $\mathcal{DD}$ invariant in the chiral 3D topological matter is distinguished by dichroic rates of optically chiral ($Z_0 \neq 0$) beams,
\beq{}
    \Delta \Gamma \equiv \Gamma_{+Z_0}  - \Gamma_{-Z_0} = \int^\infty_0 \text{d} \omega~[\Gamma_{+Z_0} (\omega) - \Gamma_{-Z_0} (\omega)],
\eeq
where the frequency ($\omega$) resolved transition rates $\Gamma_{\pm Z_0} (\omega)$ of~beams with opposite zilches $\pm Z_0$ (see Fig.~\ref{Fig:Intro}) are derived using Fermi's golden rule, as detailed in the End Matter.
\begin{figure*}
    \centering
    \includegraphics[width=\textwidth]{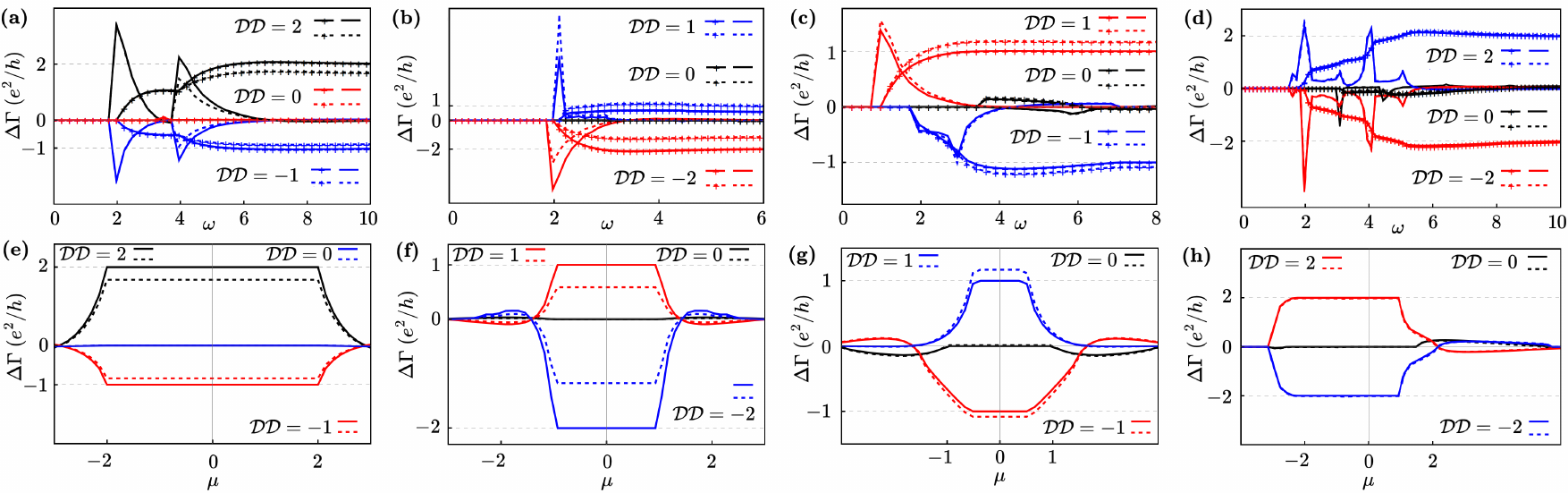}
    \caption{{\bf (a)}--{\bf (d)} TOCD of chiral topological insulators responding to light with frequency $\omega$ and zilch $Z_0$. The panels display the total gyrotropic birefringent responses (\textit{dashed}) and the topological (TOCD) contribution (\textit{bold}). The corresponding integrated responses evaluated over $(0,\omega)$ range are shown with additional markers ($+$), with quantizations converging under the excitations of full bands, in the $\omega \rightarrow \infty$ limit. {\bf (e)}--{\bf (h)} TOCD of chiral topological insulators responding to zilch $Z_0$ at different chemical potentials $\mu$. TOCD remains quantized in integers when $\mu$ lies in band gap. We show TOCD for a range of chiral topological insulator realizations: {\bf (a,e)}~Three-band chiral topological insulator~\cite{Neupert2012, Palumbo2019}. {\bf (b,f)}~Four-band chiral topological insulator~\cite{Liu2023}. {\bf (c,g)}~Two-band Hopf insulator~\cite{Moore2008, Kennedy2016}. {\bf (d,h)}~Three-band real Hopf insulator~\cite{Jankowski2024PRB}.}
    \label{Fig:Results}
\end{figure*}
As~a~central result, we find that $\mathcal{DD} \in \mathbb{Z}$ culminates in topologically quantized dichroism rate ($\Delta \Gamma$) contributions, which can be compactly written as,
\beq{}
    \Delta \Gamma = \frac{e^2 Z_0}{\hbar} \mathcal{DD}.
\eeq
It should be noted that flipping the chirality of either light $(Z_0 \rightarrow -Z_0)$ or matter $(\mathcal{DD} \rightarrow -\mathcal{DD})$ flips the TOCD rates $(\Delta \Gamma  \rightarrow -\Delta \Gamma )$. We stress that TOCD is an intrinsic 3D effect, and is fundamentally distinct from circular dichroism in recently studied 3D quantum Hall fluids that, in fact, pertain to stackings of two-dimensional topological systems~\cite{Pozo2019, Manoj2024}. Manifestly, the effect combines all three spatial dimensions, and remains invariant under the permutation of all spatial directions. The quantized effect necessitates absorption of chiral ($|Z_0|>0$) light over the frequency range $\om$ targeting all excitations between chiral bands to probe the integer $\mathcal{DD}$ invariant (see the End Matter).

We now highlight how TOCD manifests in different chiral topological insulators (Fig.~\ref{Fig:Results}). As model systems to demonstrate TOCD, we choose the hallmark three-band and four-band class~AIII insulator models~\cite{Neupert2012, Palumbo2019}. Furthermore, we show analogous responses in chiral two-band Hopf insulators~\cite{Moore2008} and three-band real Hopf insulators~\cite{Jankowski2024PRB} protected by parity-time inversion symmetry, which constitute model representatives of delicate topologies in class~A~\cite{Nelson2022, Bzdusek2025} (see also SM~\cite{SI}), showing a broader generality of TOCD beyond stable topologies in class AIII. We show how TOCD is universally manifested for a range of corresponding ${\mathcal{DD} = 0,\pm 1,\pm 2}$ invariants, which explicitly reveals an integer character of the dichroic responses. For more details on the relations of 3D chiral winding numbers and Hopf invariants and the bundle gerbe invariants integer $\mathcal{DD}$ classes, see the SM~\cite{SI}, where we further provide explicit model parameterizations for a broad range of $\mathcal{DD}$ invariants.

In Fig.~\ref{Fig:Results}(a)--Fig.~\ref{Fig:Results}(d), we show the frequency ($\omega$) resolutions of the net dichroic responses on top of partial sums converging to the quantized values of TOCD. Furthermore, we demonstrate corrections from the nontopological, dispersive terms contributing to the dichroism, which we discuss in detail in the SM. In Fig.~\ref{Fig:Results}(e)--Fig.~\ref{Fig:Results}(h), we demonstrate the changes of $\Delta \Gamma$ with chemical potential $\mu$, by setting an upper frequency limit $\omega \rightarrow \infty$. We note that TOCD is sharply quantized as long as the chemical potential $\mu$ lies in a band gap of a chiral topological insulator. Hence, we show that TOCD serves as a unique quantized optical probe of multiple Dixmier--Douady invariants in the insulating states.

\sect{Chiral topologies probed with superchiral light} We further elucidate how the integer chiral Dixmier--Douady (${\mathcal{DD} \in \mathbb{Z}}$) invariants of 3D topological matter could be experimentally accessed with superchiral light~\cite{Tang2010}. First, we briefly discuss the details of superchiral light with enhanced $Z_0$, proposed here for probing chiral topological insulators. The superchiral light with zilch $Z_0$ can be engineered with two opposite counterpropagating optical beams with light frequency $\omega$ and wavevectors $\pm\textbf{q}$, and corresponding electric fields: $\textbf{E}_1(\textbf{q}, \omega) = \vec{E}_{0,1} e^{\text{i} (\textbf{q} \cdot \textbf{r}-\omega t)}$, $\textbf{E}_2(\textbf{q}, \omega) =  \vec{E}_{0,2} e^{\text{i} (-\textbf{q} \cdot \textbf{r}-\omega t)}$, where $\vec{E}_{0,1} = |E_1|(1, \text{i})^\text{T}$, $\vec{E}_{0,2} = |E_2|(1, -\text{i})^\text{T}$ realize circular light polarizations, see Fig.~\ref{Fig:Intro}(b). The corresponding zilch $Z_0$ amounts to~\cite{Tang2010},
\beq{}
    Z_0 = \frac{\omega}{c}(E^2_1 - E^2_2),
\eeq
while the energy density $u$ at a distance $r = |\textbf{r}|$ along the superchiral beam ($\textbf{q} || \textbf{r}$) reads,
\beq{}
    u(r) = \frac{1}{2} (E^2_1 + E^2_2)-  E_1 E_2 \text{cos} (2|\textbf{q}|r).
\eeq
Correspondingly, $|Z_0|/u \gg 2|\textbf{q}|$ at $r = 0, \pi, \ldots$, as originally demonstrated in Ref.~\cite{Tang2010}. To probe responses of a slab of a topological chiral material to superchiral light, the material should be placed in the proximity of such nodal distance $r$. Notably, the magnitude of engineerable zilch $Z_0$ is only limited by the electric field strength amplitude of $E_1$ component against the $E_2$ component of the superchiral beam. 

Experimentally, the $\mathcal{DD} \in \mathbb{Z}$ invariant can therefore in principle be deduced from the difference of transmittivity of the material under the known, and controllable $Z_0$ of counterpropagating beams with circular polarizations in the superchiral light configuration, thus unraveling yet unobserved stable and delicate chiral topologies~\cite{Moore2008, Neupert2012, Kennedy2016, Alexandradinata2021, Nelson2022, Liu2023, Jankowski2025gerbe, Bzdusek2025}.

\sect{Discussion and conclusion} 
We now discuss the implications of our findings in a broader and general context of topological effects. The quantized dichroism intrinsically mixes all three spatial dimensions, and follows from the nontriviality of strong integer topological invariants underpinned by bundle gerbe cohomologies, in three spatial dimensions. Our results thus encompass a new effect with direct physical manifestations that relate to deeper mathematical structures, thereby offering the opportunity of opening a new research direction.

First, we note that the hereby retrieved optical manifestations of integer Dixmier--Douady invariants through TOCD are consistent with predictions of the topological magnetoelectric effects~\cite{Shiozaki2013, Jankowski2025gerbe} by causal Kramers--Kronig relations. We delegate the formal relation of TOCD to topological $\mathbb{Z}$-magnetoelectric phenomena to SM~\cite{SI}. In fact, the response relation between such magnetoelectric effects and TOCD can be thought of as a higher-dimensional analogue of the integer quantized Hall conductivity and quantized circular dichroism due to Chern invariants~\cite{Tran2017, Asteria2019}.

The generality of our results also allows us propose candidate materials and experimental protocols to measure this novel effect. Candidate materials for realization of the optical effect include chiral media supporting orbital magnetoelectric effects, and matter realizing $\theta$ vacua, such as axion insulators. For instance, these may involve chiral heterostructures based on the magnetic $\text{Mn} \text{Bi}_{2n} \text{Te}_{3n+1}$ class of materials. It should be noted that unlike the previously proposed protocols for measuring chiral integer topological invariants via magnetoelectric responses~\cite{Shiozaki2013}, which is sensitive to boundary conditions, our bulk optical probe relying on superchiral light remains insensitive to terminations of the chiral crystallite samples. Therefore, the proposed optical protocol may be particularly advantageous for real material searches, following the most recent developments of circular dichroism experiments in two-dimensional matter~\cite{Ghosh2024}.  

As a longer term perspective, we note that similarly to Chern insulators realizing orbital ferromagnetism, where interactions can fractionalize the value of Hall conductivity and lead to fractional quantum anomalous Hall effect, as well as to fractional quantized circular dichroism~\cite{Repellin2012}, a fractionalization of the here-introduced dichroic rates upon coupling to optical chirality could lead to fractional topological optical chirality dichroism (FTOCD) that relate to fractional $\mathcal{DD}$ invariants~\cite{Jankowski2025gerbe}. We leave the study of a possible emergence of~FTOCD for future work.

To sum up, we identify quantized dichroism in chiral topological matter, providing the first quantized optical effect intrinsic to 3D chiral magnetic topological systems including, but not restricted to, the topological class AIII of tenfold classification. We show how this can be understood as arising from higher tensor connection gauge structure and Dixmier--Douady invariants over the space of single-particle momenta. In brief, we find that the chiral topological insulating matter and higher-tensorial structures encoded in crystalline Bloch states over three spatial dimensions give rise to quantized chiroptical dichroic response due to the topological electronic chirality. Finally, our findings establish first-of-a-kind quantized dichroic response to optical zilches, therefore stimulating experimental pursuits for uncharted chiral topologies with yet underexplored physical quantity.

\sect{Acknowledgments} The authors thank Tom\'{a}\v{s}~Bzdu\v{s}ek for~illuminating discussions about mirror symmetries and Hopf bundle gerbes. W.J.J.~acknowledges funding from the Rod Smallwood Studentship at Trinity College, Cambridge. R.-J.S. acknowledges funding from an EPSRC ERC underwrite grant  EP/X025829/1 as well as Trinity College, Cambridge. This research was supported in part by grant NSF PHY-2309135 to the Kavli Institute for Theoretical Physics~(KITP).

\vspace{-12pt}

\section{End Matter}

\appendix

\sect{Details on optical chirality} We briefly elaborate on the details concerning optical chirality quantified by zilch $Z_0$. First, we show that $Z_0 = 0$ for linearly-polarized light, noting that for linear light with frequency $\omega$ and wavevector~$\vec{q}$, [$\vec{E}(\textbf{q},\omega) = \vec{E}_0 e^{\text{i} (\textbf{q} \cdot \textbf{r}-\omega t)}$]: 
\begin{align}
Z_0 &= \vec{E} \cdot (\nabla \times \vec{E}) + \vec{B} \cdot (\nabla \times \vec{B}) = \vec{E} \cdot (\textbf{q} \times \vec{E}) + \vec{B} \cdot (\textbf{q} \times \vec{B}) \\&= \textbf{q} \cdot (\vec{E} \times \vec{E}) + \textbf{q} \cdot (\vec{B} \times \vec{B}) = 0,
\end{align}

using the cyclic property of the triple scalar product. For~circularly-polarized light, $\textbf{E}(\textbf{q},\omega) = \frac{|\textbf{E}|}{\sqrt{2}} (1,\pm \text{i})^\text{T} e^{\text{i}(\textbf{q} \cdot \textbf{r} - \omega t)}$, the zilch analogously amounts to $Z_0 = \pm\frac{\omega|\textbf{E}|^2}{c}$~\cite{Tang2010}. Thus, the chirality of light depends on both the electric field magnitude $|\textbf{E}|$ and the frequency of radiation $\omega$.

\sect{Derivation of the topological optical chirality dichroism} We hereby derive topological dichroism in response to optical chirality (TOCD), directly from Fermi's golden rule (FGR),
\beq{}
    \Gamma(\omega) = \frac{2\pi}{\hbar} \sum_{n,m} f_{nm} |\bra{\psi_{m}} \Delta H \ket{\psi_{n}}|^2 \delta (E_n - E_m + \hbar \om),
\eeq
with time-dependent perturbation electric and magnetic fields $E_\mu(t), B_\mu(t)$: $\Delta H = e \hat{r}^
\mu E_\mu - \hat{M}^\mu B_\mu $. Here, we consider the position operator $\hat{r}^
\mu$ ($\mu = x,y,z$) under a minimal coupling in length gauge, orbital magnetization $\hat{M}^\mu 
 \equiv \frac{d \hat{G}^\mu}{dt}$, and excitations between states $\ket{\psi_n}$, $\ket{\psi_m}$ with energies $E_n, E_m$ and occupation probabilities $f_n, f_m$ given by Fermi-Dirac distribution. $f_{nm} = f_n - f_m$ denote occupation differences. From now on, we consider single-particle Bloch eigenstates, $\ket{\psi_n} \rightarrow \ket{\psi_{n\kv}}$, with momenta $\textbf{k}$ and energy eigenvalues $E_n \rightarrow E_{n\kv}$. 
 
 Upon introducing a perturbation that involves a single Fourier component $\omega$ with electromagnetic fields $\textbf{E}(\om), \textbf{B}(\om)$ in $\Delta H$, we obtain,
\begin{widetext}
\beq{}
    \Gamma (\om) = \frac{2\pi}{\hbar} \sum_{\kv} \Big[ e^2 |\textbf{E}(\om)|^2 \sum_{n,m} f_{nm} Q^{mn}_{\mu \nu} + e E_\mu (\om) B_\nu(\om) \sum_{n,m} f_{nm} \bra{\psi_{m\kv}} \hat{r}^\mu \ket{\psi_{n\kv}} \bra{\psi_{n\kv}} \hat{M}^\nu \ket{\psi_{m\kv}} + \text{c.c.} + O(\textbf{B}^2) \Big] \delta (\om_{mn} - \om),
\eeq
\end{widetext}
with $\omega_{mn} \equiv \frac{E_{m\kv} - E_{n\kv}}{\hbar}$, multiband quantum geometric \mbox{tensor}, ${Q^{mn}_{\mu \nu} = \bra{\psi_{m\kv}} \hat{r}^\mu \ket{\psi_{n\kv}} \bra{\psi_{n\kv}} \hat{r}^\nu \ket{\psi_{m\kv}} = A^{mn}_\mu  A^{nm}_\nu}$~\cite{Ahn2020, Ahn2021, Jain2025}, and~non-Abelian Berry connection, $A^{mn}_\mu = \text{i} \braket{u_{m\kv} | \partial_{k_\mu} u_{n\textbf{k}}}$, defined in terms of Bloch states, $\ket{\psi_{n\kv}} = e^{\text{i} \kv \cdot \textbf{r}} \ket{u_{n\kv}}$. Following \mbox{Faraday's} Law: ${\nabla \times \vec{E} = - \dot{\vec{B}}}$, $\partial_{[\mu} E_{\nu]}(\om) = -\text{i}\om \varepsilon^{\mu \nu \rho} B_{\rho}(\om)$, with antisymmetrization $[ \ldots ]$, upon using the definitional relation, $\hat{M}^\mu = \text{i}\om \hat{G}^\mu = \text{i} \om \varepsilon^{\mu \nu \rho} \hat{G}_{\nu \rho} $, we obtain,
\begin{widetext}
\beq{}
    \Gamma (\om) = \frac{2\pi}{\hbar} \sum_{\kv} \Big[ e^2 |\textbf{E}(\om)|^2 \sum_{n,m} f_{nm} Q^{mn}_{\mu \nu} + e E_\mu(\om) \partial_\nu E_{\rho}(\om) \sum_{n,m} f_{nm} A^{nm}_\mu  G^{mn}_{\nu \rho} + \text{c.c.} + O(\textbf{B}^2) \Big] \delta (\om_{mn} - \om),
\eeq
\end{widetext}
where $G^{mn}_{\nu \rho} = \bra{\psi_{m\kv}} \hat{G}_{\nu \rho} \ket{\psi_{n\kv}}$. In particular, the first term provides for the circular dichroism given by the Berry curvature~\cite{Tran2017, Asteria2019} and vanishes in band insulators with trivial Chern numbers, as considered in this work. From now on, we focus on the second term. The geometric contribution to interband  moment $G^{mn}_{\nu \rho}$ reads~\cite{Malashevich2010, Trifunovic2019},
\begin{align}
    G^{mn}_{\nu \rho} = -e\sum_{p \neq m} (A_\nu^{mp} A_\rho^{pn} - A_\rho^{mp} A_\nu^{pn}),
\end{align}
which sums over virtual transitions through states $p$. We further note that the geometric interband moments $G^{mn}_{\nu \rho}$ introduce the band torsion~\cite{Ahn2021, Jankowski2024PRL},
\begin{align}
    \mathcal{T}^{nm}_{\mu \nu \rho} = -(1/e) A^{nm}_{\mu} G^{mn}_{\nu \rho} = A^{nm}_\mu \sum_{p} (A_\nu^{mp} A_\rho^{pn} - A_\rho^{mp} A_\nu^{pn}), 
\end{align}
in the FGR. Moreover, following Ref.~\cite{Jankowski2025gerbe}, we define,
\begin{align}
    \mathcal{H}^{nm}_{\mu \nu \rho} =  -2\mathcal{T}^{nm}_{[\mu \nu \rho]},
\end{align}
and hence $\Delta \Gamma = \int^\infty_0 \dd \om~[\Gamma_{+Z_0}(\om)-\Gamma_{-Z_0}(\om)]$ yields,
\beq{}
    \Delta \Gamma = \frac{2\pi e^2}{\hbar} [\vec{E} \cdot (\nabla \times \vec{E})] \sum_{\kv} \sum_{n,m} f_{nm} \mathcal{H}^{nm}_{\mu \nu \rho}.
\eeq
Upon setting $f_n =1$, $f_m =0$, in the zero temperature limit,
\begin{align}
    \Delta \Gamma &= \frac{2\pi e^2}{\hbar} [\vec{E} \cdot (\nabla \times \vec{E})] \sum_{\kv} \sum_{n,m} \mathcal{H}^{nm}_{\mu \nu \rho} \nonumber  \\&= \frac{ e^2}{\hbar} [\vec{E} \cdot (\nabla \times \vec{E})] \sum_{n,m} \mathcal{DD}_{nm} \\&= \frac{ e^2}{\hbar} [\vec{E} \cdot (\nabla \times \vec{E})] \mathcal{DD} \nonumber,
\end{align}
where $\sum_\kv = \int_\text{BZ}\frac{\text{d}^3 \kv}{(2\pi)^3}$ sums over the Brillouin zone (BZ) of~particle momenta, and we trace over the $\mathcal{DD}_{nm}$ invariants of~individual bands ($\mathcal{DD} = \sum_{n,m} \mathcal{DD}_{nm}$). Finally, upon combining with the definition of $Z_0 \sim \vec{E} \cdot (\nabla \times \vec{E})$, and assuming that the magnetic coupling term effects are negligible~\cite{Tang2010}, we~finally obtain, 
\begin{align}
    \Delta \Gamma = \frac{e^2 Z_0}{\hbar}  \mathcal{DD} \nonumber,
\end{align}
i.e., the final TOCD formula capturing the topological \mbox{response} to zilch, as central to the main text.

\bibliography{references}
\end{document}


\title{{\textsc{supplemental material}} \\  Topological Optical Chirality Dichroism}

\newcommand{\TCM}{{TCM Group, Cavendish Laboratory, University of Cambridge, J.\,J.\,Thomson Avenue, Cambridge CB3 0HE, United Kingdom}}
\newcommand{\MCR}{Department of Physics and Astronomy, University of Manchester, Oxford Road, Manchester M13 9PL, United Kingdom}
\newcommand{\Coimbra}{{Department of Physics, University of Coimbra, Rua Larga, 3004-516 Coimbra, Portugal}}
\newcommand{\Dublin}{{School of Theoretical Physics, Dublin Institute for Advanced Studies,
10 Burlington Road, Dublin D04 C932, Ireland}}


\author{Wojciech J. Jankowski}
\email{wjj25@cam.ac.uk}
\affiliation{\TCM}

\author{Giandomenico Palumbo}
\email{giandomenico.palumbo@gmail.com}
\affiliation{\Coimbra}
\affiliation{\Dublin}

\author{Robert-Jan Slager}
\email{robert-jan.slager@manchester.ac.uk}
\affiliation{\MCR}
\affiliation{\TCM}

\date{\today}

\maketitle

\begin{widetext}

\tableofcontents 

\newpage

\section{Bundle gerbes and models}
We provide further mathematical details on bundle gerbe constructions and characteristics of the models with arbitrarily high $\mathcal{DD}$ invariants. We begin with reviewing the constructions of tensor Berry connections over bundle gerbes and more formal details on the gerbe invariants, following Ref.~\cite{Jankowski2025gerbe}. As mentioned in  the main text, the tensor connection amounts to ${B^{nm}_{ij} = \phi_{nm}F^{nm}_{ij}}$, with  $F^{nm}_{ij} = \partial_{k_i} A^{nm}_j -  \partial_{k_j} A^{nm}_i$, the Berry flux, consistent with the notation of~\cite{Jankowski2025gerbe}.

The pseudoscalar field $\phi(\kv)$, with $\textbf{k} = (k_x,k_y,k_z)$ a single-particle Bloch state momentum, can be constructed from non-Abelian Berry connection $\vec{A}_{nm}(\kv) = \text{i} \bra{u_{n\kv}} \ket{\nabla_\kv u_{m\kv}}$ as follows~\cite{Jankowski2025gerbe},
\beq{}
    \phi_{nm}(\kv) = \int^\kv_0 \dd \kv' \cdot \vec{A}_{nm}(\kv'),
\eeq
where the line integral is evaluated along the shortest path within the parameter space. Accordingly, the higher-tensorial Berry curvature flux three-form reads,
\beq{}
    \mathcal{H}^{nm}_{xyz} = \partial_{k_x} B^{nm}_{yz} + \partial_{k_y} B^{nm}_{zx} + \partial_{k_z} B^{nm}_{xy}.
\eeq
The corresponding Dixmier--Douady gerbe invariants read~\cite{Palumbo2019, Palumbo2021, Jankowski2025gerbe},
\beq{}
    \mathcal{DD}_{nm} = -\frac{1}{4\pi^2} \int_\text{BZ} \dd^3 \kv~ \mathcal{H}^{nm}_{xyz}.
\eeq
The $\mathcal{DD} \in \mathbb{Z}$ invariants correspond to the nontrivial cohomology ring elements of $H^3 (S^3, \mathbb{Z}) \cong \mathbb{Z}$~\cite{Murray1996}. In the following, we~focus on distinct realizations of Dixmier--Douady bundle gerbe invariants, which can be accessed through the dichroic response demonstrated in the main text. In particular, these include three-band and four-band chiral topological insulators in~class AIII, two-band Hopf insulators, and three-band real Hopf insulators.

\begin{figure*}[t!]
    \centering
    \includegraphics[width=0.8\textwidth]{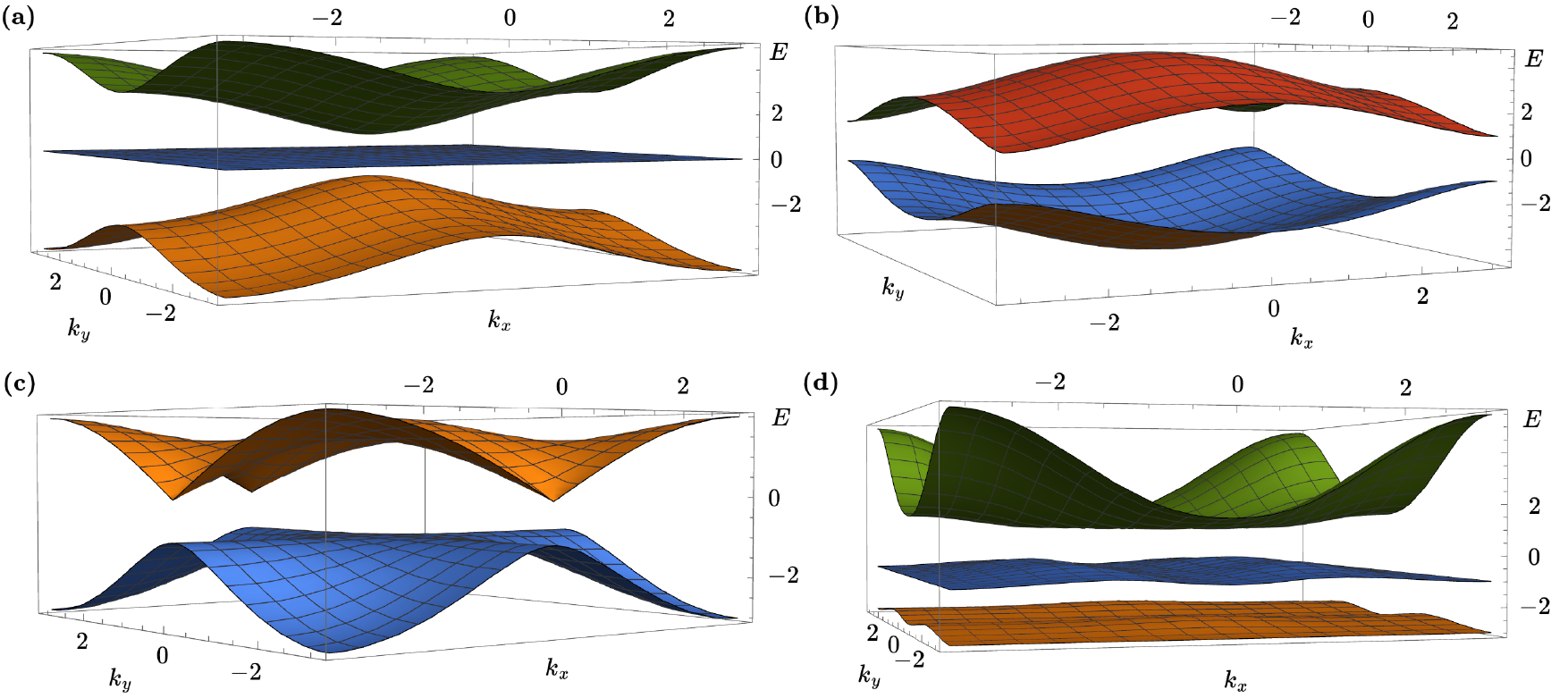}
    \caption{Band structures of distinct chiral topological insulator realizations: {\bf (a)}~Three-band chiral topological insulator (${k_z = 0, m = 4}$)~\cite{Neupert2012, Palumbo2019}. {\bf (b)}~Four-band chiral topological insulator (${k_z = 1, m = 2}$)~\cite{Liu2023}. {\bf (c)}~Two-band Hopf insulator (${k_z = 0, m = 1}$)~\cite{Moore2008, Kennedy2016}. {\bf (d)}~Three-band real Hopf insulator (${k_z = 0, m = 1}$)~\cite{Jankowski2024PRB}.}
    \label{Fig:S1}
\end{figure*}
\subsection{Three-band chiral insulators}

The Hamiltonian of a three-band chiral insulator studied in the main text reads~\cite{Neupert2012},
\beq{}
    H^\text{chiral}_{(3)} = \begin{pmatrix}
    0 & 0 & \sin k_x - \text{i} \sin k_y\\
    0 & 0 & \sin k_z - \text{i}[m - \cos wk_x - \cos k_y - \cos k_z] \\
    \sin k_x + \text{i} \sin k_y  &  \sin k_z + \text{i}[m - \cos wk_x - \cos k_y - \cos k_z]  & 0  \\
    \end{pmatrix}.
\eeq
We show the bands of the corresponding three-band chiral insulator in Fig.~\ref{Fig:S1}(a).

The Hamiltonian  $H^\text{chiral}_{(3)}$ realizes $\mathcal{DD}_{11} = 2w$ for $0 < |m| < 3/2$, $\mathcal{DD}_{11} = -w$ for $3/2 < |m| < 3$, and $\mathcal{DD}_{11} = 0$ for $3 < |m|$.

\subsection{Four-band chiral insulators}

The four-band chiral insulator Hamiltonian studied in the main text reads~\cite{Liu2023},
\beq{}
   H^\text{chiral}_{(4)} = \begin{pmatrix}
    0 & 0 & \sin k_y - \text{i} \sin k_z & m + e^{-\text{i} w k_x} + \cos k_y + \cos k_z\\
    0 & 0 & m + e^{\text{i} w k_x} + \cos k_y + \cos k_z  & - \sin k_y - \text{i} \sin k_z \\
    \sin k_y + \text{i} \sin k_z  &  m + e^{-\text{i} w k_x}  + \cos k_y + \cos k_z  & 0 & 0 \\
    m + e^{\text{i} w k_x} + \cos k_y + \cos k_z  & - \sin k_y + \text{i} \sin k_z & 0 & 0 \\
    \end{pmatrix}.
\eeq
We show the doubly-degenerate bands of the corresponding four-band chiral insulator in Fig.~\ref{Fig:S1}(b).

The Hamiltonian  $H^\text{chiral}_{(4)}$ realizes $\mathcal{DD}_{11} = -2w$ for $0 < |m| < 3/2$, $\mathcal{DD}_{11} = w$ for $3/2 < |m| < 3$, and $\mathcal{DD}_{11} = 0$ for $3 < |m|$.

\subsection{Two-band Hopf insulators}

The two-band Hopf insulator Hamiltonian studied in the main text reads~\cite{Moore2008, Kennedy2016},
\beq{}
 H^\text{Hopf}_{(2)} = \begin{pmatrix}
    m - \cos w k_x - \cos k_y & \cos k_z  \sin w k_x + \sin k_y \sin k_z -\text{i}[\sin k_x \sin k_z -\cos k_z \sin k_y]\\
   \cos k_z \sin w k_x + \sin k_y \sin k_z +\text{i}[\sin k_x \sin k_z -\cos k_z \sin k_y] &  -m + \cos w k_x + \cos k_y \\
    \end{pmatrix},
\eeq
which yields $\mathcal{DD}_{11} = -w$ for $0 < m < 2$, $\mathcal{DD}_{11} = w$ for $-2 < m < 0$, and $\mathcal{DD}_{11} = 0$ for $2 < |m|$.

We show the bands of the corresponding two-band Hopf insulator in Fig.~\ref{Fig:S1}(d).

\subsection{Three-band real Hopf insulators}

The Hamiltonian of the three-band real Hopf insulator studied in the main text compactly reads~\cite{Jankowski2024PRB}
\beq{}
 H^\text{Hopf}_{(3)} = \begin{pmatrix}
-3 + \frac{1}{2}d_x^2
&
\frac{1}{2}d_y
d_x
&
\frac{1}{2}d_z
d_x
\\[6pt]

\frac{1}{2}d_y
d_x
&
-1 + \frac{1}{2}d_y^2
&
\frac{1}{2}d_z
d_y
\\[6pt]

\frac{1}{2}d_z
d_x
&
\frac{1}{2}d_z
d_y
&
1 + \frac{1}{2}d_z^2
\end{pmatrix},
\eeq
where we adapt a shorthand notation,
\beq{}
   d_x = \left(\cos k_z\sin k_x - \sin k_y\sin k_z\right),
\eeq
\beq{}
   d_y = \left(\cos k_z\sin k_y + \sin k_x\sin k_z\right),
\eeq
\beq{}
   d_z = \left(m - \cos k_x - \cos k_y\right).
\eeq
We show the bands of the corresponding three-band real Hopf insulator in Fig.~\ref{Fig:S1}(d).

The Hamiltonian  $H^\text{Hopf}_{(3)}$ realizes $\mathcal{DD}_{12} = -2$ for $0 < m < 2$, $\mathcal{DD}_{12} = 2$ for $-2 < m < 0$, and $\mathcal{DD}_{12} = 0$ for $2 < |m|$.

Notably, in three-band real Hopf insulator models~\cite{Jankowski2024PRB}, under the resolution of identity, the interband $\mathcal{DD}_{nm}$ invariants (${n,m = 1,2,3}$) are not independent:

 \begin{align}
    \mathcal{DD}_{12} &= -\frac{1}{4\pi^2} \int_\text{BZ} \dd^3 \kv~ \mathcal{H}^{12}_{xyz} = -\frac{1}{4\pi^2} \int_\text{BZ} \dd^3 \kv~ \textbf{A}_{12} \cdot (\nabla \times \textbf{A}_{12}) \nonumber\\&= \frac{1}{4\pi^2} \int_\text{BZ} \dd^3 \kv~ \epsilon^{ijk} A_i^{12} A_j^{13} A_k^{32} = \frac{1}{4\pi^2} \int_\text{BZ} \dd^3 \kv~\epsilon^{ijk} A_j^{13} A_k^{32} A_i^{21} \\&= -\frac{1}{4\pi^2} \int_\text{BZ} \dd^3 \kv~ \textbf{A}_{13} \cdot (\nabla \times \textbf{A}_{13}) = -\frac{1}{4\pi^2} \int_\text{BZ} \dd^3 \kv~ \mathcal{H}^{13}_{xyz} =  \mathcal{DD}_{13}\nonumber.
\end{align}
Upon retrieving analogous relations for the other combinations of bands, we have:
\beq{}
    \mathcal{DD}_{12} = \mathcal{DD}_{21} = \mathcal{DD}_{13} = \mathcal{DD}_{31} = \mathcal{DD}_{23} = \mathcal{DD}_{32}.
\eeq

\section{Relations to magnetoelectric effects} Beyond the Fermi's golden rule (FGR) derivation of the quantization included in the End Matter, we furthermore connect the spectral decomposition of the response tensor central to the quantized response to the static magnetoelectric polarizability. The static magnetoelectric polarizability is defined as,
\beq{}
    \alpha_{ij}(0) = \frac{\partial M_j}{\partial E_i}
\eeq
where $M_j$ represents a bulk magnetization induced by the electric field $E_i$, with $i,j = x,y,z$. To connect TOCD to magnetoelectric responses, we begin by decomposing the optical magnetoelectric polarizability $\alpha_{ij}(\om)$ in the static limit $\om \rightarrow 0$, following Ref.~\cite{Ahn2022}, which obtains,
\beq{}
    \alpha_{ij}(0) = \alpha^\text{K}_{ij}(0) + \alpha^\text{top}_{ij}(0)
\eeq
and determines the ground state magnetoelectric response. $\alpha^\text{K}_{ij}(0)$ is the Kubo term, and $\alpha^\text{top}_{ij}(0)$ is a topological term, which amongst other terms, includes the Chern-Simons angle $\theta$~\cite{Ahn2022}, $\alpha^\text{top}_{ij}(0) = \frac{e^2}{\hbar c} \theta + \alpha^\text{QM}_{ij} (0) + \alpha^\text{MD}_{ij} (0)$, where $\alpha^\text{QM}_{ij} (0)$ denotes a~quadrupolar term, and $ \alpha^\text{MD}_{ij} (0)$ is a metric dipole term absent in the insulating state, as identified in Ref.~\cite{Ahn2022}. Using Kramers--Kronig relations for the topological term, with $\mathcal{P}$ denoting a principal value, we have,
\beq{}
\text{Re}~\alpha^{\text{top}}_{ij}(0) = \frac{1}{\pi} \mathcal{P}~ \int^\infty_0 \dd\om' \frac{\text{Im}~\alpha^\text{top}_{ij}(\om')}{\om' - \om}
\eeq
where the optical magnetoelectric polarizability tensor, in the notation of TOCD matrix elements, reads~\cite{Ahn2022},
\beq{}
    \alpha^{\text{top}}_{ij}(\omega) = \frac{e}{\hbar} \int_{\text{BZ}} \text{d}^3 \kv~\sum_{m,n} \frac{\om^2_{mn}}{\om^2_{mn}-\om^2} f_{nm} \epsilon^{jkl}~A_i^{nm} G_{kl}^{mn}.
\eeq
Here, $G^{nm}_{kl}$ is an interband matrix elements given by the geometric contribution to orbital magnetization, as defined in the End Matter, while the non-Abelian Berry connection $A_i^{nm}$ amounts to the interband polarization captured by electric transition dipole matrix elements $r^{nm}_i = \bra{\psi_{n\kv}} \hat{r}_i \ket{\psi_{m\kv}} = (1-\delta_{mn}) A_i^{nm}$. On contracting a resolution of identity and utilizing Sokhotski--Plemelj theorem, we end up with,
\begin{align}
   \text{Tr}~\alpha^\text{top}_{ij}(0) = \int_{\text{BZ}} \dd^3 \kv~ \sum_{n,m} f_{nm} \epsilon^{ikl} \text{Im}~A^{nm}_i G^{mn}_{kl} \nonumber \\= \frac{e^2}{hc}\int_{\text{BZ}} \dd^3 \kv~ \sum_{n,m} f_{nm}  \mathcal{H}^{nm}_{xyz} = \frac{e^2}{\hbar c} \sum_{n,m} \mathcal{DD}_{nm},
\end{align}
which takes values in integers $(\mathcal{DD}_{nm} \in \mathbb{Z})$ in chiral topological insulators~\cite{Shiozaki2013} and Hopf insulators~\cite{Jankowski2025gerbe}. For the class~AIII Hamiltonians, the $\mathcal{DD}$ invariants coincide with the chiral winding numbers~\cite{Shiozaki2013, Palumbo2019}. The emergence of the $\mathcal{DD} \in \mathbb{Z}$ invariant in~$\text{Tr}~\alpha^\text{top}_{ij}(0)$ underpins topological $\mathbb{Z}$-magnetoelectric effects, hereby retrieved via Kramers--Kronig relations, consistently with the construction and other central findings of Ref.~\cite{Jankowski2025gerbe}.

On the other hand, the nontopological optical correction to gyrotropic birefringence~\cite{Ahn2022}, present beyond TOCD, amounts to, 
\begin{align}
    \alpha^{\text{non-top}}_{ij}(\om) =& \frac{e^2}{2\hbar} \int_\text{BZ}  \frac{\dd^3 \kv}{(2 \pi)^3} \sum_{n,m} \frac{f_{nm}}{E_{m\kv}-E_{n\kv} - \hbar \omega} \Big[r^{nm}_i \sum_{p \neq m} (r^{mp}_k v^{pn}_j - r^{mp}_j v^{pn}_k) \Big],
\end{align}
where the velocity operator matrix elements $v^{pn}_j = \langle \psi_{p\kv} | \hat{v}_j \ket{\psi_{n\kv}}$, and the velocity operator $\hat{v}_j = \frac{\text{d} \hat{r}_j}{\text{d} t} = \frac{\text{i}}{\hbar}[\hat{H}, \hat{r}_j]$. Under Kramers--Kronig relations, $\alpha^{\text{non-top}}_{ij}(\om)$ can be connected to the nontopological part of the static magnetoelectric tensor, which reads~\cite{Malashevich2010},
\begin{align}
    \alpha^{\text{non-top}}_{ij}(0) =& \frac{e^2}{\hbar c} \int_\text{BZ} \frac{\dd^3 \kv}{(2 \pi)^3} \sum_{n,m} \text{Im} \Bigg(\varepsilon^{jlq}  \frac{\bra{\partial_{k_q} u_{n\kv}} \partial_{k_l} (H + E_{n\kv}) \big|u_{m\kv} \Big\rangle r^{mn}_i}{E_{m\kv}-E_{n\kv}} \Bigg).
\end{align}
Unlike the purely geometric and energy-independent topological term, the nontopological correction to optical magnetoelectric polarizability heavily depends on the band dispersion, as internally captured by the matrix elements of the velocity operator.

\end{widetext}

\clearpage

\bibliographystyle{apsrev4-1}
\bibliography{references.bib}